\documentclass[conference]{IEEEtran}
\IEEEoverridecommandlockouts
\usepackage{cite}
\usepackage{amsmath,amssymb,amsfonts}
\usepackage{algorithmic}
\usepackage{subcaption}
\usepackage{graphicx}
\usepackage{textcomp}
\usepackage{xcolor}
\usepackage{tikz}
\usepackage{standalone} 
\usepackage{tikzscale} 
\usetikzlibrary{backgrounds,fit,positioning}
\def\BibTeX{{\rm B\kern-.05em{\sc i\kern-.025em b}\kern-.08em
    T\kern-.1667em\lower.7ex\hbox{E}\kern-.125emX}}
\begin{document}

\title{Symbiotic Message Passing Model for Transfer Learning between Anti-Fungal and Anti-Bacterial Domains
\thanks{$^*$Corresponding author. yoni.savir@technion.ac.il
}
}

\author{\IEEEauthorblockN{Ronen Taub}
\IEEEauthorblockA{\textit{Dept. of Physiology,} \\
\textit{Biophysics \& Systems Biology} \\
\textit{Faculty of Medicine} \\ 
\textit{Technion}\\
Haifa, Israel \\
} 

\and 
\IEEEauthorblockN{Tanya Wasserman}
\IEEEauthorblockA{\textit{Dept. of Physiology,} \\
\textit{Biophysics \& Systems Biology} \\
\textit{Faculty of Medicine} \\ 
\textit{Technion}\\
Haifa, Israel \\
} 

\and
\IEEEauthorblockN{Yonatan Savir$^*$}
\IEEEauthorblockA{\textit{Dept. of Physiology,} \\
\textit{Biophysics \& Systems Biology} \\
\textit{Faculty of Medicine} \\ 
\textit{Technion}\\
Haifa, Israel \\
} 
}

\maketitle

\begin{abstract}
Machine learning, and representation learning in particular, has the potential to facilitate drug discovery by screening billions of compounds. For example, a successful approach is representing the molecules as a graph and utilizing graph neural networks (GNN). Yet, these approaches still require experimental measurements of thousands of compounds to construct a proper training set. While in some domains it is easier to acquire experimental data, in others it might be more limited. For example, it is easier to test the compounds on bacteria than perform in-vivo experiments. Thus, a key question is how to utilize information from a large available dataset together with a small subset of compounds where both domains are measured to predict compounds' effect on the second, experimentally less available domain. Current transfer learning approaches for drug discovery, including training of pre-trained modules or meta-learning, have limited success. In this work, we develop a novel method, named Symbiotic Message Passing Neural Network (SMPNN), for merging graph-neural-network models from different domains. Using routing new message passing lanes between them, our approach resolves some of the potential conflicts between the different domains, and implicit constraints induced by the larger datasets. By collecting public data and performing additional high-throughput experiments, we demonstrate the advantage of our approach by predicting anti-fungal activity from anti-bacterial activity. We compare our method to the standard transfer learning approach and show that SMPNN provided better and less variable performances. Our approach is general and can be used to facilitate information transfer between any two domains such as different organisms, different organelles, or different environments. \\

\end{abstract}


\section{Introduction}
One of the promising future trajectories of drug discovery is harnessing the power of machine learning to screen for billions of compounds in silico \cite{a12,a13,a14,a15,a16}. The main advantage of this approach is that it requires a reasonable experimental effort to produce a training set based only on the order of thousands of compounds. One of the key challenges in applying machine learning for drug discovery is the representation of the molecule. The classic approach is to define a feature vector composed of properties based on chemical properties, such as competition of functional groups \cite{a17,a18,a19}. Once the molecule is represented as a feature vector, the machine can learn which features are associated with some function. In this approach, the definition of the features representing the molecule requires extensive knowledge and prior knowledge in biochemistry. Representation learning, a new approach in which the machine learns not only what are the features that are relevant to a function but also learns the relevant features that represent the data has emerged as a powerful tool \cite{a20}. 

A successful approach for representing molecules is to represent them as a graph and utilize graph neural networks (GNN) such as graph convolutional network (GCN) \cite{a21,a22,a23} and directed-message passing neural network (D-MPNN) \cite{a24,a25}. It has several variations and implementations for drug discovery. In practice, these methods encode the spatial information of the atomic structure graph into the atoms' hidden state, resulting in fixed-size encodings. These encodings are then fed into a binary classifier, to produce the final prediction. 

\begin{figure*}
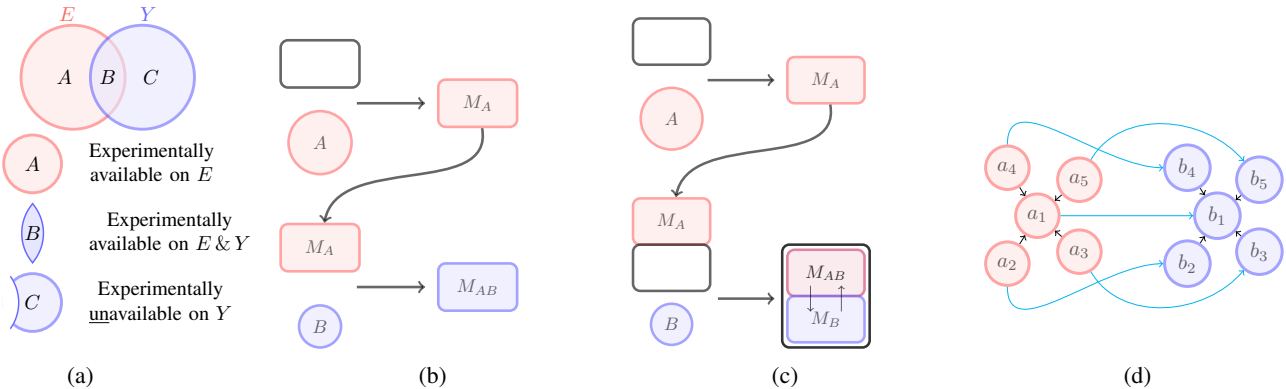

    \centering
     \begin{subfigure}[b]{0.225\textwidth}
        \includegraphics[width=1.1\linewidth]{figures/figure1a.tex}
        \caption{}
        \label{fig:1a}
     \end{subfigure}
     \hfill
     \begin{subfigure}[b]{0.225\textwidth}
        \includegraphics[width=0.8\linewidth]{figures/figure1b.tex}
        \caption{}
        \label{fig:1b}
     \end{subfigure}
     \hfill
     \begin{subfigure}[b]{0.225\textwidth}
        \includegraphics[width=0.8\linewidth]{figures/figure1c.tex}
        \caption{}
        \label{fig:1c}
     \end{subfigure}
     \hfill
     \begin{subfigure}[b]{0.225\textwidth}
        \includegraphics[width=1\linewidth]{figures/figure1d.tex}
        \caption{}
        \label{fig:1d}
     \end{subfigure}
        \caption{(a) Details the settings of the targeted transfer learning problem. $E$ and $Y$ denote two domains to which we want to provide predictions, e.g. two organisms, two organs, two environments, etc. In our case, we focus on bacteria and yeast as the two domains. Group $A$ marks the available compounds, while group $C$ marks the compounds we wish to infer from $A$ with the help of $B$. $B$ is the intersection where the properties are known for both $E$ and $Y$. (b)  The dual-step transfer-learning routine. Where an untrained model is optimized for observations from $A$, and then optimized again for observations from $B$. The model's architecture does not vary between the steps in a significant manner. Subfigures (c) and (d) illustrate our method. In the second optimization step, we added a new model instance, which communicates with the first, converged model $M_A$ creating a symbiosis. The manner in which the two models communicate is illustrated in Subfigure (d). Which is a message-passing scheme, operating at the atoms-level.}
        \label{fig:my_label}
\end{figure*}

One of the main challenges of applying the machine learning approach is the need for a training set that usually contains $10^3-10^4$ compound measurements. While in some domains we have high availability of data, resulting in diverse, rich training sets in others it might be more limited due to, for example, experimental challenges. For example, it is much easier to measure the effect of compounds on microorganisms in high-throughput compared to cell-line measurements, or \textit{in-vivo} experiments. Thus, a key question is to what degree can we utilize information from a large available dataset together with a small subset of compounds where we have both domains to predict compounds' effect on the second, less experimentally available domain (Fig. \ref{fig:1a}).  These domains are not limited to two different organisms but could be relevant when considering different organs, different environments, different ages, etc.
 
A possible solution for this challenge is  Transfer Learning \cite{b8}. In transfer learning, the model is first optimized on the other-domain dataset. And then, in a secondary optimization process, it is optimized over the available data from the relevant domain. Modified versions of the default transfer-learning scheme, have been shown to work for drug discovery, resulting in improved predictions \cite{b9,b10,b15}. Another common transfer-learning scheme is  Meta-Learning \cite{b10}, where only a subset of the other-domain data is used, leaving out data samples that do not contain mutual information or create conflicts during secondary training.

In this work, we propose a novel method to transfer information from one GNN model to another. The emphasis in our method is on the symbiosis between the fused model-parts, rather than on the data-sources. Hence, it can also be combined with the Meta-Learning framework. Our underlying idea is to add new trainable components to the model in the secondary optimization, in a unique manner. Hence, the new components are independent of the other domain raw data. By doing so, we relax the implicit restriction to the environment of the initial condition, derived by the first optimization in the other domain. Meaning, the model can infer new rules, which are unrelated to the first domain. 

We demonstrate how our method can facilitate the prediction of antifungal activity, the effect of compounds on the viability of the budding yeast \textit{S.cerevisiae}, from antibacterial activity on the bacteria \textit{E.coli}. Both \textit{E.coli} and \textit{S.cerevisiae} are common model organisms for a variety of conditions. While \textit{E.coli} is a prokaryote, \textit{S.cerevisiae} is a eukaryote. Therefore, it provides a good example of two domains in which the interplay between compounds' effects on each domain is not trivial. We compared the different transfer-learning methods and show that our method provides better results in term of AUROC and that it is closer to Pareto optimality in terms of F1-score. 

\section{Methods and Training Data}

\subsection{Training Data}

For this paper, we have utilized two datasets. The first dataset is a public one \cite{b7}, containing the effect of 2336 FDA-approved compounds on the bacteria \textit{E.coli}. From this collection, 120 compounds exhibited antibiotic activity. The second dataset was produced in our lab. We have used a similar FDA-approved compounds library containing 1467 compounds. The number of shared compounds, between the two libraries, is 824. We measured the response of \textit{S.cerevisiae} by incubating them with the drugs and monitoring their growth. The yeast were added to each well with a final $OD_{600}$ of $0.05$. Initial $OD_{600}$ was measured using the Infinite M200 Pro plate reader (Tecan), after which the plates were sealed with an AeraSealTM (Excel Scientific) and incubated in a Multitron Pro shaker-incubator (Infors HT) at 30°C, 1000 rpm and 70\% humidity. Final $OD_{600}$ was measured after $16h$. We have defined compounds as having an anti-fungal activity if they reduced the yeast growth significantly, compared with control growth, p-cal$<$0.05. In total, there are 43 positively-labeled compounds with anti-fungal activity. Hereafter, for simplicity, we will refer to \textit{E.coli} by the term 'bacteria', and to \textit{S.cerevisiae} by the term 'yeast'.


\subsection{Message Passing Neural Network}

Message Passing Neural Network (MPNN) is a term for GNN classifier architecture, that consists of a GNN encoder, aggregation operator, and predictor functional \cite{b7}. In our work, we followed the implementation available in the code environment, distributed in the Chemprop package (which is the code environment for \cite{b7} and available on GitHub). According to this implementation, the graph encoder operation can be presented by the following system:
\begin{equation}\label{eq:init}
     \vec{m}_{i\rightarrow j}^{(0)} = ReLU \left( W_I \cdot \begin{pmatrix}
     \begin{pmatrix} \vec{a}_j \end{pmatrix} \\ 
     \begin{pmatrix} \vec{b}_{i\rightarrow j}\end{pmatrix} 
    \end{pmatrix} \right) \quad if \quad \vec{b}_{i\rightarrow j} \neq 0
\end{equation}
\begin{equation}\label{eq:recursive}
     \vec{m}^{(t+1)}_{j\rightarrow i} = ReLU \left( \vec{m}^{(0)}_{j\rightarrow i} + W_H \cdot \sum_{k=1}^{N_A} \begin{pmatrix}
     \begin{pmatrix}  \vec{m}^{(t)}_{k \rightarrow j}  \end{pmatrix} \\ 
      \begin{pmatrix} \vec{b}_{k \rightarrow j} \end{pmatrix} 
    \end{pmatrix} \right) 
\end{equation} 
\begin{equation}\label{eq:sum}
     \vec{e}_i = \sum_{j=1}^{N_a} \vec{m}^{(T)}_{j \rightarrow i} 
\end{equation}
\begin{equation}\label{eq:out}
\vec{\alpha}_i =  ReLU \left( W_O \cdot \begin{pmatrix}
  \begin{pmatrix} \vec{e}_i \end{pmatrix} \\
 \begin{pmatrix} \vec{a}_i \end{pmatrix} 
\end{pmatrix} \right) 
\end{equation}

where $\vec{a}_i \in \mathrm{R}^{d_a}$ is the feature vector of the $i^{th}$ atom, containing the relevant chemical properties for each atom. Similarly, $\vec{b}_{i\rightarrow j} \in \mathrm{R}^{d_b}$ is the feature vector of the bond going from the $i^{th}$ atom to the $j^{th}$ atom. Equation (eqn. \eqref{eq:init}) is the initialization step, projecting the atoms features and connected bonds onto the hidden feature space. Where $W_I \in \mathrm{R}^{(d_a+d_b) \times d_h}$ is the projection matrix. Equation (eqn. \eqref{eq:recursive}) represent the recursive encoding operator. In each iteration $t$, for each bond, the states of the neighbor atoms (of the bond's source-atom) are summed and linearly transformed with $W_h\in \mathrm{R}^{d_h \times d_h}$ (in order to update the corresponding bond-message). After $T$ iterations, all the resulted messages are summed in (eqn. \eqref{eq:sum}), to produce encodings at the atoms-level. Finally, $\alpha_i$ is the final encoding of the $i^{th}$ atom after the external stimulation, described in (eqn. \eqref{eq:out}).

\subsection{Transfer Learning}

Transfer Learning is a methodology to transfer gathered intelligence, implicitly contained within one model, to another model. When having two groups of datasets, the common approach for transfer learning is a dual-step procedure, where the model is first trained on the knowledge of larger group $A$ and then trained again on the partial knowledge of group $B$ (Fig. \ref{fig:1b}). We used this approach to benchmark our method. In our case, we first trained on the bacteria dataset and than re-trained the model using different sets with different number of compounds, that are shared between the bacteria and yeast dataset. The size of the sets goes from 100 to 824 in jumps of 100. 

\subsection{SMPNN: Symbiotic Message Passing Neural Network} 

The motivation for the SMPNN model is to create a beneficial symbiosis between model-parts trained on data from group $A$, and model-parts trained on available data from group $B$. This way, we have unique model parameters supporting the domain knowledge in group $A$ (i.e., optimized on), unique model parameters supporting the domain knowledge in group $B$, and unique model parameters that support both. While in regular transfer-learning methods and or meta-learning, we have only the latter. Hence, the hybrid model provides the maximum flexibility to fit observed data (\ref{fig:1c}) 

Our basic model instance carries the operation described by equations (eqn. \eqref{eq:init} - \eqref{eq:out}). The MPNN architecture contains a graph-encoder, an aggregation operator, and a fully-connected network (as for binary classifier). After the aggregation operator, the observability of the encoder state is limited, since it combines all the atoms' encodings. Before the aggregation operator and past the encoder, different converged MPNN instances produce different features for each atom, creating a synchronization problem. Therefore, we decided to connect two different MPNN pipelines, at the encoder level.


In Fig. \ref{fig:1d}, there is an illustration of the manner in which we connect two GNN encoders. We route the messages from each atom in the enriched model to its mirror atom in the less-enriched model. We have decided to not pass the messages directly, but rather linearly project them with projection matrix $W_S\in \mathrm{R}^{d_h\times d_h}$. The reasoning behind this is to allow the second pipeline to maintain a different hidden-state space (feature space) from the first pipeline converged space. The modified versions of equations (eqn. \eqref{eq:init} - \eqref{eq:out}) equal:

\begin{equation}\label{eq:init2}
     \widetilde{\vec{m}}_{i\rightarrow j}^{(0)} = ReLU \left( \widetilde{W}_I \cdot \begin{pmatrix}
     \begin{pmatrix} \vec{a}_j \end{pmatrix} \\ 
     \begin{pmatrix} \vec{b}_{i\rightarrow j}\end{pmatrix} 
    \end{pmatrix} \right) \quad if \quad \vec{b}_{i\rightarrow j} \neq 0
\end{equation}
\begin{equation}\label{eq:init3}
     \vec{m}_{i\rightarrow j}^{(0)} = ReLU \left( W_I \cdot \begin{pmatrix}
     \begin{pmatrix} \vec{a}_j \end{pmatrix} \\ 
     \begin{pmatrix} \vec{b}_{i\rightarrow j}\end{pmatrix} 
    \end{pmatrix} \right) \quad if \quad \vec{b}_{i\rightarrow j} \neq 0
\end{equation}
\begin{equation}\label{eq:recursive2}
     \widetilde{\vec{m}}^{(t+1)}_{j\rightarrow i} = ReLU \left( \widetilde{\vec{m}}^{(0)}_{j\rightarrow i} + \widetilde{W}_H \cdot \sum_{k=1}^{N_A} \begin{pmatrix}
     \begin{pmatrix}  \widetilde{\vec{m}}^{(t)}_{k \rightarrow j}  \end{pmatrix} \\ 
      \begin{pmatrix} \vec{b}_{k \rightarrow j} \end{pmatrix} 
    \end{pmatrix} \right) 
\end{equation} 
\begin{align}\label{eq:recursive3}
     m^{(t+1)}_{j\rightarrow i} = ReLU \large \Bigg( \normalsize m^{(0)}_{j\rightarrow i} + W_H \cdot \sum_{k=1}^{N_A} \begin{pmatrix}
     \begin{pmatrix}  \vec{m}^{(t)}_{k \rightarrow j}  \end{pmatrix} \\ 
      \begin{pmatrix} \vec{b}_{k \rightarrow j} \end{pmatrix} 
    \end{pmatrix}  \nonumber \\
    + W_S \cdot \widetilde{W}_H \cdot \sum_{k=1}^{N_A} \begin{pmatrix}
     \begin{pmatrix}  \widetilde{\vec{m}}^{(t)}_{k \rightarrow j}  \end{pmatrix} \\ 
      \begin{pmatrix} \vec{b}_{k \rightarrow j} \end{pmatrix} \end{pmatrix} \large \Bigg) \normalsize
\end{align}
\begin{equation}\label{eq:sum2}
     \vec{e}_i = \sum_{j=1}^{N_a} \vec{m}^{(T)}_{j \rightarrow i} 
\end{equation}
\begin{equation}\label{eq:out2}
\vec{\alpha}_i =  ReLU \left( W_O \cdot \begin{pmatrix}
  \begin{pmatrix} \vec{e}_i \end{pmatrix} \\
 \begin{pmatrix} \vec{a}_i \end{pmatrix} 
\end{pmatrix} \right) 
\end{equation}

where $\widetilde{W}_I,\widetilde{W}_H$ are the weights from the converged model in the 'transfer-from' domain (i.e., where the training data is available). The matrices $W_I,W_H,W_O$ are the new trainable component added to assemble the hybrid model. The equations (eqn. \eqref{eq:init2},\eqref{eq:recursive2}) maintain the original message passing scheme of the 'transfer-from' model. While equation (eqn. \eqref{eq:recursive3}), is where the transferring occurs, with matrix $W_S$. 

To summarize the entire method, we detail the full-training process with the following recipe:
\begin{enumerate}
    \item Train MPNN instance over the 'transfer-from' enriched data according to the original labels. 
    \item Construct SMPNN instance according to equations (eqn. \eqref{eq:init2} - \eqref{eq:out2}).
    \item Initialize the weights of the fully-connected network binary classifier of SMPNN, to be the same ones as in the converged MPNN. 
    \item Initialize the weight matrices $\widetilde{W}_I,\widetilde{W}_H$ to be $W_I,W_H$ from converged MPNN.
    \item Train the SMPNN over the available subset of examples from the 'transfer-to' targeted domain. 
    
\end{enumerate}

\section{Results}

In this work, we focused on evaluating the effectiveness of transfer learning between the bacteria domain $E$ and the yeast domain $Y$. To do this, we divided the data populations into 3 groups: $E, E\cap Y, Y/\{E\cap Y\}$. Group $A$, the 'transfer-from' population contains the compounds in $E$ with labels that denote their antibacterial activity. Group B, which is the training set for the second optimization has the compounds from $E\cap Y$ with antifungal activity. Test group C has the compounds in  $Y/\{E\cap Y\}\}$. In numbers, $|E|=2336$ compounds, $|E\cap Y|=824$ compounds, $|Y/\{E\cap Y\}|=643$ compounds. To test the effect of adding more compounds we used intersection subsets with increasing numbers. $B_{N}$ is a subset of $B$ that contains $N$ examples from $B$. We examined various $B_N$ groups for various values of $N$, to observe in what manner the transfer-learning improves with more data available for the second training. More specifically, we randomly selected $B_N$ out of the intersection population (at each time), with the condition $B_N \subset B_{N+1}$ (for the same random-generator-seed).

\begin{figure}[h]
    \centering
    \includegraphics[width=0.49\textwidth]{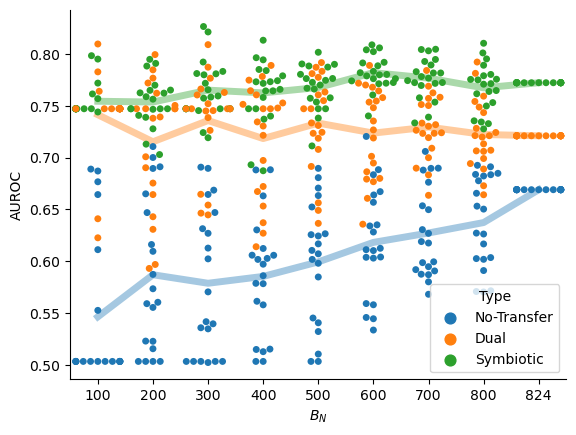}
    \caption{The effect of transfer learning dataset size on the AUROC (area under the ROC curve). A model was trained only on the compounds with known effects on bacteria ($A$ in Fig. \ref{fig:1a}) and then was retrained using a subset of compounds to which both the bacteria and yeast effect are known ($A\cap B$ in Fig. \ref{fig:1a}). The AUROCs are the result of predicting antifungal activity on unseen compounds ($C$ in Fig. \ref{fig:1a}). The x-axis is T the number of  compounds in retrain set $B_N$.  'No-Transfer' denotes training only occurs only on $B_N$ (without the initial training on set $A$). 'Dual' and 'Symbiotic' are the two examined transfer-learning methods. Each dot represents a different random choice of the $B_N$ compounds. The lines are the averages.}
    \label{fig:results_a}
\end{figure}

First, to estimate the interplay between the bacteria domain $E$ and the yeast domain $Y$, we trained the model only with the bacteria ground-truth labels and tested on the yeast set, that is training on $A$ and testing on $C$. The AUROC (area under the ROC curve) received for this model is $0.73$. To make sure that this results is indeed related to the information encapsulated within the bacteria ground-truth label and not the result of training on the large chemical space, we tested the results of training on a similar number of compounds but with random ground-truth labels. We randomized the labels 20 times to get a representative population. The distribution of the AUROC in this case is in the range of $0.58\sim 0.63$. This indicate that the antibacterial data indeed carries information for the antifungal activity.

Next, tested the effect of using transfer-learning, that is using also ground-truth information from the yeast measurements, and evaluate the improvement delta when using SMPNN over the common transfer-learning (\ref{fig:1b}). For this we compared SMPNN with two alternatives: training only on $B_N$ (i.e., not using bacteria domain knowledge, 'No-Transfer'), and training on $A$ first and then training again on $B_N$ for the same model (i.e., 'Dual-Training'). We gradually changed the value of $N$ to observe the manner in which the performance transition. The results are presented in Fig. \ref{fig:results_a}. One can see that for every value of $N$, it applies that the AUROC statistics of SMPNN is better than the AUROC statistics for 'Dual-Training', and the AUROC statistics for 'Dual-Training' is better than the AUROC statistics when not performing transfer-learning at all. Interestingly, the AUROC of SMPNN is always greater than $M_A$, but that is not the case always for 'Dual-Training'. This implies, that there might be conflicts during the second training, when using the same model. Another interesting fact, is that the AUROC for SMPNN in the case of $B_{100}$ is better than the results for 'Dual-Training' and no-transfer-learning in the case of $B_{900}$. 

\begin{figure}[h]
     \centering
    \includegraphics[width=0.49\textwidth]{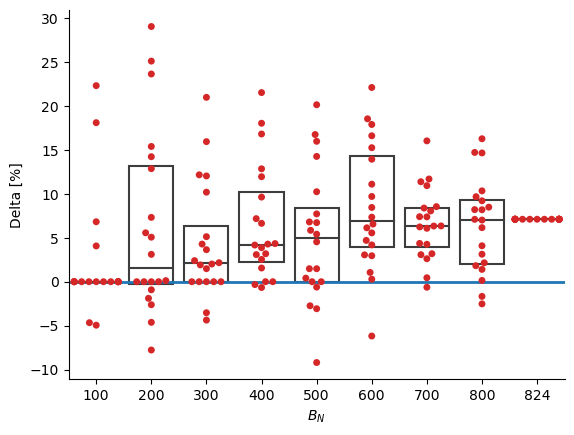}
    \caption{Swarmplot with boxplot of improvement delta in AUROC between the symbiotic model (i.e. SMPNN) and the traditional dual-training scheme. Each dot represents a different random choice of the $B_N$.}
    \label{fig:results_ba}
\end{figure}

 Fig. \ref{fig:results_ba} illustrated the improvement of using in SMPNN in terms of AUROC. The improvement delta increases with $N$. Meaning, the marginal value of additional data, is greater with the Symbiotic model. Hence, the results for $B_{800}$ should contain the best improvement delta. In numbers, the improvement delta when using SMPNN is $8\%$ on average with respect to 'Dual-Training', and $33\%$ on average with respect to not performing transferring learning at all (for $B_{800}$). Besides the improvement median performance, using SMPNN yields a more reproducible cross validation. There is a $58\%$ decrease decrease in the SMPNN standard-deviation compared with the  'Dual-Training' method, and $69\%$ decrease compared with no transfer learning. 

\begin{figure}[h]
    \centering
    \includegraphics[width=0.49\textwidth]{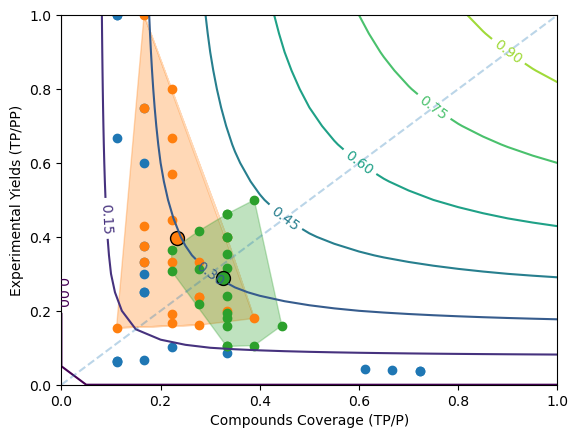}
    \caption{Scatter-plot of the  $B_{800}$ models, comparing 'No-transfer-learning' (blue), 'Dual-Training' (orange), and 'Symbiotic' model (green). The precision and recall values are derived from the predictions of the converged model (in each point) with respect to the same test, i.e., group $C$. The contour lines are constant F1 score values. The average point for each method is the large circle with black-color edges. The blue dashed line is precision equals to recall.}
    \label{fig:results_b}
\end{figure}

To analyze the results further, we examined the F1-score metric on the precision-recall plane. We took the results for $B_{800}$ in the three methods. As a reminder, $B_{800}$ means that we train on the bacteria dataset (group A) first. Then we train on the randomly picked 800 molecules from group $B$, named as $B_{800}$. And finally, test on group $C$ (which are molecules from the yeast dataset that do not appear in the intersection). The precision and recall results are displayed in Fig. \ref{fig:results_b}. The fact that the area covered by 'Dual-Training' is significantly larger than the area for 'Symbiotic' model. That is due to the fact that the performance standard deviation in the case of the 'Symbiotic' model is significantly lower. It is insightful to consider the resolution to the tradeoff between precision and recall.  The SMPNN model also provides a more balanced solution to the recall-precision tradeoff. In the case there is a tradeoff between recall and sensitivity such that both of them have the same cost, then the SMPNN model is closer to Pareto optimality.

\section{Conclusions}

While previous works in the field of drug discovery, have focused more on the correct mixture of datasets (either manually or with meta-learning) for fruitful transfer-learning, our Symbiotic model (SMPNN) puts the focal point on the manner in which domain models are fused together. To emulate the meta-learning scheme, we reconstructed the sub-optimality domain (meta-learning operates on) by randomizing $B_N$. The results indicate that SMPNN excels for all the values of $N$ (in almost all instances of $B_N$). We hypothesize that this is due to the fact that our specific symbiosis allows the model to learn rules that are not necessarily connected to, or agree with, the first converged model. This is supported also by the fact that the improvement delta values (from results) increase with $N$. In general, in the case of transfer learning, reproducibility of the  results, as we mix datasets from different domains, which are not explicitly linked, is crucial.  Our method exhibits significant improvement in the standard deviation of the performances.

 The aim of any drug-discovery model is to provide predictions that have to be validated experimentally. It is insightful to consider the performance of drug discovery in terms of compound coverage (recall) and experimental yield (precision). The number of predicated positive compounds determines the cost of the experimental validation. Hence, the experimental yield, which is the ratio between true positive and predicted positive, is desired to receive a high value (closer to 1). The second important metric is compound coverage, which is the ratio between true positive and overall positive compounds. This metric is also desired to be a high number, close to 1. Meaning, we wish to discover new drugs, and not receive the same already recorded drugs. There is usually a built-in trade-off between these two metrics. A natural metric that accounts for both of these quantities is the F1-score. Our results indicate that, if recall and precision have a tradeoff between them such that they have similar costs, the Symbiotic model is closer to the Pareto optimal resolution.

While we have demonstrated the value of our approach using two different organisms,  our approach is general and can be used to facilitate information transfer between any two domains.  

\section*{Acknowledgment}
This work is supported by the Israeli Ministry of Science and Technology (MOST) grant \#2149.

\end{document}